\documentclass[preprint,12pt]{elsarticle}

\usepackage{amssymb,amsmath,amsthm,bm}

\biboptions{sort&compress}

\begin{document}

\begin{frontmatter}

%% Title, authors and addresses

%% use the tnoteref command within \title for footnotes;
%% use the tnotetext command for the associated footnote;
%% use the fnref command within \author or \address for footnotes;
%% use the fntext command for the associated footnote;
%% use the corref command within \author for corresponding author footnotes;
%% use the cortext command for the associated footnote;
%% use the ead command for the email address,
%% and the form \ead[url] for the home page:
%%
%% \title{Title\tnoteref{label1}}
%% \tnotetext[label1]{}
%% \author{Name\corref{cor1}\fnref{label2}}
%% \ead{email address}
%% \ead[url]{home page}
%% \fntext[label2]{}
%% \cortext[cor1]{}
%% \address{Address\fnref{label3}}
%% \fntext[label3]{}

\title{New astrophysical limit on neutrino millicharge}

%% use optional labels to link authors explicitly to addresses:
%% \author[label1,label2]{<author name>}
%% \address[label1]{<address>}
%% \address[label2]{<address>}

\author[1,2]{Alexander I. Studenikin}
\ead{studenik@srd.sinp.msu.ru}
\author[1]{Ilya V. Tokarev}
\ead{tokarev.ilya.msu@gmail.com}

\address[1]{Department of Theoretical Physics, Faculty of Physics, Moscow State University, Moscow 119991, Russia}
\address[2]{Joint Institute for Nuclear Research, Dubna 141980, Moscow Region, Russia}

\begin{abstract}
An impact of a nonzero neutrino millicharge in astrophysics is tested. It is shown that in astrophysical environments electromagnetic interactions of the neutrino millicharge with strong electromagnetic fields as well as weak interactions of the neutrinos with dense background matter can produce new phenomena accessible
for astrophysical observations. On this basis a new limit on the neutrino millicharge $q_0<1.3\times10^{-19}e_0$ is obtained. This limit is among the strongest astrophysical constraint on the neutrino millicharge. Some other possible applications of the obtained results to astrophysics are discussed in details.
\end{abstract}

\begin{keyword}
%% keywords here, in the form: keyword \sep keyword

%% MSC codes here, in the form: \MSC code \sep code
%% or \MSC[2008] code \sep code (2000 is the default)

\end{keyword}

\end{frontmatter}

%%
%% Start line numbering here if you want
%%
% \linenumbers

%% main text

\section{Introduction}

Studies of nontrivial neutrino electromagnetic properties~\cite{Giunti:2008ve,Broggini:2012df}
is one of the important
issues of the nowadays physics beyond the Standard Model~\cite{Studenikin:2008bd,Studenikin:2013yaa,Studenikin:2013saa,Giunti:2014ixa}.
The updated review on the present status of neutrino electromagnetic properties is given in \cite{Giunti:2014ixa}. In this short paper we deal with a nonzero electric millicharge.

It is usually believed that the neutrino electric charge is zero. In the Standard Model neutrinos are massless and electrically neutral particles. However, the electric charge of neutrinos as massive particles is an controversial question that can open a window to ``new physics''.

There are a lot of limits on the neutrino millicharge known in literature~\cite{Giunti:2014ixa,Raffelt:1996wa}. The best model independent astrophysical constraint $q_0\leq2\times10^{-17} e_0$ was obtained from the analysis of the neutrino signal from SN1987A~\cite{Barbiellini:1987zz} (here $q_0$ and $e_0$ are the absolute values of neutrino and electron charges respectively). The most severe constraint $q_0\lesssim3\times10^{-21} e_0$ was obtained from the consideration of charge conservation law in neutron beta decay and direct measurements of the neutrality of matter~\cite{Marinelli:1983nd} and neutron itself~\cite{Baumann:1988ue}. Also note a new approach to the problem that is based on the analysis of results of reactor neutrino magnetic moment experiments and yields the limit at the level of a few $10^{-12} e_0$~\cite{Studenikin:2013my}.

In turn, recently we have focused on new astrophysical effects originated by
both weak and electromagnetic interactions of millicharged neutrinos with background environment.
For description of neutrinos propagation in a dense rotating matter we use the
method of exact solutions of the modified Dirac equation for the neutrino wave function \cite{Studenikin:2008qk}.
In particular, a new mechanism of a star rotation frequency shift due to neutrinos escaping the star (termed ``Neutrino Star Turning'' mechanism, $\nu S T$) is predicted~\cite{Studenikin:2012vi}.
We have obtained a new limit on the neutrino millicharge using the estimation
 of the proposed $\nu ST$ mechanism impact on the dynamics of core-collapse of a supernova. In addition, possible applications of this and other predicted effects have been discussed in details.

\section{Millicharged neutrino in dense magnetized rotating matter}

In order to predict new astrophysical effects and phenomena originated by the
millicharged neutrinos in astrophysical environment one should to describe
the neutrino behavior in such extreme background conditions.

We use in our studies of neutrinos under extreme external conditions, in particular
in strong electromagnetic fields and dense matter,
the method of exact solutions of modified Dirac equations~\cite{Studenikin:2008qk}.
Within this method, a millicharged neutrino quantum states in an external magnetic field
and dense background matter is described by the modified Dirac equation
\begin{equation}
\label{Dirac}
\left(\gamma_{\mu}(p^{\mu}+q_0A^{\mu})-\frac12\gamma_{\mu}(1+\gamma_5)f^{\mu}-m\right)\Psi(x)=0,
\end{equation}
where $A^{\mu}$ and $f^{\mu}$ are electromagnetic and matter potentials respectively (it is supposed that the neutrino millicharge is negative). In case of the neutrino propagation in the magnetized rotating matter where both matter rotation vector $\bm\omega$ and the magnetic field $\bm B$ are coincided with the third coordinate axis $\bm e_z$ the solution of Eq.~(\ref{Dirac}) was obtained in the following form~\cite{Studenikin:2012vi,Studenikin:2013yaa,Studenikin:2013saa}
\begin{equation}
\label{spectr}
p_0=\sqrt{p_3^2+2N|q_0B-2Gn\omega|+m^2}+Gn,
\end{equation}
where $G=\frac{G_F}{\sqrt{2}}$ ($G_F$ is the Fermi constant), $\omega$ is a matter angular  velocity, $n$ is a matter number density and $N=0,1,2..$ is a discrete number that numerates the modified Landau levels. The energy spectrum~(\ref{spectr}) of the millicharged neutrino in the dense magnetized rotating matter is quantized due to both electromagnetic interactions of the neutrino millicharge with the constant magnetic field and weak interactions of the neutrino with the dense rotating matter.

Note that the neutrino quantum states that also account for the electromagnetic interaction of a neutrino magnetic moment with the external magnetic field were obtained in~ \cite{Balantsev:2012ep,Balantsev:2013aya,Studenikin:2012vi}.

Within a quasi-classical interpretation the neutrino discrete energy states~(\ref{spectr}) can be explained as a result of action of the effective force \cite{Studenikin:2008qk,Studenikin:2012vi}
\begin{equation}
\label{force}
\bm F=(q_0B-2Gn\omega)\left[\bm{e}_z\times\bm\beta\right],
\end{equation}
where $\bm\beta$ is a neutrino velocity. Note that the force~(\ref{force})
is of different nature that one of the classical Lorentz force
because of the presence of the impact from weak interactions with particles of the background.

\section{Astrophysical applications}

The force~(\ref{force}) seems to be very weak for any reasonable choice of background parameters, but nevertheless, can produce new astrophysical effects that can be observed in terrestrial experiments~\cite{Studenikin:2012vi}. In particular, during a supernova core collapse due to the action of the force~(\ref{force}) escaping neutrinos can be deflected on an angle
\begin{equation}
\Delta\phi \simeq \frac{R_S}{R}\sin\theta, \quad R=\sqrt{\frac{2N}{|q_0B-2Gn\omega|}},
\end{equation}
where $R_S$ is the radius of the star, $R$ is the radius of the neutrino trajectory and $\theta$ is an azimuthal angle of neutrino propagation. Thus, we predict that initially coincided light and neutrino beams will be spatial separated after passing through the dense rotating magnetized matter. Therefore, in terrestrial experiments joint observations of initially coincided light and neutrino signals from astrophysical transient sources should not occur due to their spatial separation $\Delta L\simeq\Delta\phi L$ ($L$ is distance to the sources). This new effect can explain recent results of the ANTARES experiment~\cite{Adrian-Martinez:2014wzf}.

On the other hand the feedback of the effective force~(\ref{force}) from the escaping neutrinos to the star should effect the star evolution. In particular, a torque produced by the escaping neutrinos shifts the star angular velocity
\begin{equation}
\label{delta_omega}
|\triangle\omega_0|=\frac{5N_{\nu}}{6M_{S}}|q_0B-2Gn\omega_0|,
\end{equation}
where $\triangle\omega_0=\omega-\omega_0$ ($\omega_0$ is an initial star angular velocity), $M_S$ is the star mass and $N_{\nu}$ is the number of the escaping neutrinos. We have termed the phenomenon as the ``Neutrino Star Turning'' ($\nu ST$) mechanism~\cite{Studenikin:2012vi}. Note that depending on the neutrino millicharge sign the star rotation due to the $\nu ST$ mechanism can either spin up ($\triangle\omega_0>0$ for $q_{\nu}<0$) or spin down ($\triangle\omega_0<0$ for $q_{\nu}>0$).

In case of zero neutrino millicharge the $\nu ST$ mechanism is produced only due to the weak interactions and yields
\begin{equation}
\label{delta_omega_weak}
\frac{|\triangle\omega_0|}{\omega_0}\simeq10^{-8}.
\end{equation}
where we have considered the star with mass $M_S = 1.4  M_{\odot}$ ($M_{\odot}$ is the Solar mass) and $N_{\nu}=10^{58}$ escaped neutrinos with energy $\sim10$ MeV \cite{Hirata:1987hu}.

The value of the relative rotation frequency shift~(\ref{delta_omega_weak}) is very close to a
sporadic increase of a pulsar rotation frequency (a pulsar glitch, see, for instance,
\cite{Rosen:2012ty}).
The obtained results are also important in light of the recently observed ``anti-glitch'' event \cite{Archibald:2013kla} that is a sudden decrease of a pulsar rotation frequency. The $\nu ST$ mechanism can be used to explain both glitches and ``anti-glitches'' as well.

\section{New astrophysical limit on neutrino millicharge}

To obtain a new limit on the neutrino millicharge we have estimated the impact of the electromagnetic part of the $\nu ST$ mechanism on the dynamics of pulsar formation in a supernova explosion.

From Eq.~(\ref{delta_omega}) we obtain
\begin{equation}
\label{delta_omega_nonzero_charge}
\frac{|\triangle\omega_0|}{\omega_0}=\frac{7.6q_0}{e_0}\times
10^{18}\left(\frac{P_0}{10\text{ s}}\right)
\left(\frac{N_{\nu}}{10^{58}}\right)
\left(\frac{1.4  M_{\odot}}{M_{S}}\right)
\left(\frac{B}{10^{14}\textrm{G}}\right),
\end{equation}
where $P_0$ is a pulsar initial spin period.

All estimations of feasible initial pulsars rotation periods give the values that are very close to the present observed periods. Therefore, possible existence of a nonzero negative neutrino millicharge should not significantly speed up the rotation of a born pulsar (or speed down in case of a positive neutrino millicharge). From the straightforward demand $|\triangle\omega_0| < \omega_0$ and Eq.~(\ref{delta_omega_nonzero_charge}) we have obtained the upper limit on the neutrino millicharge~\cite{Studenikin:2012vi}
\begin{equation}
\label{bound_q_nu}
q_0<1.3\times10^{-19} e_0.
\end{equation}

That is, in fact, one of the most severe astrophysical limits on the neutrino millicharge \cite{Giunti:2014ixa,Raffelt:1996wa}.

\section{Acknowledgments}
One of the authors (A.S.) is thankful to Arcadi Santamar\'{\i}a, Salvador Mart\'{\i} and Juan Fuster for the kind invitation to participate at the ICHEP 2014 conference and to all of the organizers for their hospitality in Valencia. The work on this paper has been partially supported by the Russian Foundation for Basic Research (grants No. 14-02-31816 and 14-22-03043).

%% The Appendices part is started with the command \appendix;
%% appendix sections are then done as normal sections
%% \appendix

%% \section{}
%% \label{}

%% References
%%
%% Following citation commands can be used in the body text:
%% Usage of \cite is as follows:
%%   \cite{key}         ==>>  [#]
%%   \cite[chap. 2]{key} ==>> [#, chap. 2]
%%

%% References with BibTeX database:
%%\nocite{*}
\bibliographystyle{model1a-num-names}
\bibliography{new_limit}

%% Authors are advised to use a BibTeX database file for their reference list.
%% The provided style file elsarticle-num.bst formats references in the required Procedia style

%% For references without a BibTeX database:

% \begin{thebibliography}{00}

%% \bibitem must have the following form:
%%   \bibitem{key}...
%%

% \bibitem{}

% \end{thebibliography}

\end{document}